\documentclass{aa}
\usepackage{psfig}

\newcommand{\SAXI}{SAX\,J1324.5$-$6313}
\newcommand{\SAXII}{SAX\,J1752.3$-$3128}
\newcommand{\SAXIII}{SAX\,J1753.5$-$2349}
\newcommand{\SAXIV}{SAX\,J1806.5$-$2215}
\newcommand{\SAXV}{SAX\,J1818.7$+$1424}
\newcommand{\GRS}{GRS\,1741.9$-$2853}

\newcommand{\nh}{N_{\rm H}}
\newcommand{\cmsq}{{\rm atoms\,cm}^{-2}}
\newcommand{\ergs}{{\rm erg\,s}^{-1}}

\newcommand{\ergcms}{{\rm erg\,cm}^{-2}{\rm s}^{-1}}
\newcommand{\ltap}{\mathrel{\hbox{\rlap{\lower.55ex \hbox {$\sim$}}
                   \kern-.3em \raise.4ex \hbox{$<$}}}}
\newcommand{\gtap}{\mathrel{\hbox{\rlap{\lower.55ex \hbox {$\sim$}}
                   \kern-.3em \raise.4ex \hbox{$>$}}}}

\begin{document}


\title{Chandra follow-up of bursters with low persistent emission }

\author{R.\ Cornelisse\inst{1,2} \and  F.\ Verbunt\inst{2} \and
        J.J.M.\ in 't Zand\inst{2,1} \and E.\ Kuulkers\inst{1,2} 
        \and J.\ Heise\inst{1}}
\offprints{R.\ Cornelisse}
\mail{R.Cornelisse@sron.nl}

\institute{SRON National Institute for Space Research, Sorbonnelaan 2, 
              3584 CA Utrecht, The Netherlands  
         \and  Astronomical Institute, Utrecht University,
              P.O.Box 80000, 3508 TA Utrecht, The Netherlands}

\date{\today / Accepted date}   
 
\abstract{We report on Chandra ACIS-S observations of five type\,I
X-ray bursters with low persistent emission: \SAXI, \SAXII, \SAXIII,
\SAXIV, and \SAXV. We designate candidate persistent sources for
four X-ray bursters. All candidates are detected at a
persistent luminosity level of 10$^{32-33}$ $\ergs$, comparable to
soft X-ray transients in quiescence. From the number of bursters with
low persistent emission detected so far with the Wide Field Cameras,
we estimate a total of such sources in our Galaxy between 30 and 4000.  
}

\maketitle

\section{Introduction}

Many low-mass X-ray binaries show bursts of X-rays which are characterized
by a rapid rise and exponential decay, and by a black body spectrum with
spectral softening during the decay i.e.\ the emitter cools. Such type\,I
X-ray bursts are interpreted as thermonuclear flashes on surfaces of
neutron stars, and thus effectively identify the emitting source
as a neutron star as opposed to a black hole.
The theory of these bursts predicts a relation between the accretion
rate onto the neutron star, as measured by the persistent X-ray luminosity,
and the properties of the X-ray burst. Briefly, for very low and very
high accretion rates, no X-ray bursts are expected, because thermonuclear
fusion is steady (Fujimoto et~al. 1987). At intermediate accretion rates, 
hydrogen/helium fusion occurs sporadically in bursts, and the burst frequency 
is a function of the accretion rate per square meter on the neutron star.
Because the effectively accreting area of the neutron star is also a
function of the accretion rate, the burst frequency is a non-monotonic
function of the persistent X-ray luminosity. Recent reviews
of burst theory are given by Bildsten (1998, 2000).

Low-mass X-ray binaries are discovered as either persistent sources or
transient sources. The transient sources with neutron stars show
outbursts lasting for weeks, sometimes up to years, at luminosities
above $10^{36}$ $\ergs$. During their quiescent state their luminosity
drops to a level of $10^{32-33}$ $\ergs$ (e.g. Campana
et~al. 1998), and the time averaged luminosities are $\ltap10^{36}$
$\ergs$ (e.g. White et~al. 1984). Most bursts are emitted by
systems at luminosities $\gtap10^{36}$ $\ergs$, e.g.\ the transients
Aql\,X-1 and Cen\,X-4 emitted X-ray bursts when they were in outburst
(Koyama et~al. 1981, Matsuoka et~al. 1980).

The Wide Field Cameras (WFC) on board the Italian-Dutch Satellite
BeppoSAX discovered sporadic type\,I bursts from nine previously
unknown burst sources, which had persistent X-ray fluxes below the WFC
detection limit of a few times $10^{-10}$ $\ergcms$ (2-28 keV).
At 8 kpc, the distance of the Galactic center, these flux limits
correspond to luminosities of $\sim10^{36}$ $\ergs$. Four
of the nine previously unknown burst sources were detected with other 
instruments at fluxes well below the WFC detection limit
(see Table\,\ref{previous}). The five other bursters are listed 
in Table\,\ref{overzicht}. In this article we present Chandra
observations which we obtained in order to determine the flux
levels of these five burst sources.

The persistent luminosities of the nine previously unknown burst
sources are (possibly far) below $10^{36}$ $\ergs$, i.e.\ below the
level X-ray bursts are usually observed. This is the reason why Cocchi
et~al.\ (2001) suggested that these sources are members of a new class
of bursters with low persistent emission (see also Cornelisse
et~al. 2002).

\begin{table}[t]
\caption{Overview of the detection of four of the low persistent emission 
bursters. For each source we list the instrument which detected the
source, the date of observation and the persistent flux in
$10^{-11}$ $\ergcms$ plus passband in keV.
References: a. Kaptein et~al. 2000, b. Cocchi et~al. 1999, c. Pavlinsky 
et~al. 1994,  d. Cornelisse et~al. 2002, f. Antonelli et~al. 1999, g. in 't 
Zand et~al. 2002 (in preparation).
\label{previous}}
\begin{tabular}{l@{\hspace{0.20cm}}c@{\hspace{0.20cm}}c@{\hspace{0.20cm}}c@{\hspace{0.15cm}}c@{\hspace{0.15cm}}c@{\hspace{0.15cm}}}
\hline
source & instrument & date & $F$ & range &ref\\
\hline
1RXS\,J1718.4$-$4029 & ROSAT/P & 1990 & 1    & 2-10 &a\\
1RXS\,J1718.4$-$4029 & ROSAT/H  & 1994 & 0.4  & 2-10 &a\\
\GRS                 & GRANAT     & 1990 & 19   & 4-30 &b,c\\
SAX\,J1828.5$-$1037  & ROSAT/P & 1993 & 0.19 & 0.5-2.5 &d\\
SAX\,J2224.9$+$5421$^e$ & SAX/NFI & 1999 & 0.013 & 2-10&f,g\\   
\hline
\multicolumn{6}{l}{$^e$ Observation a few hours after burst.}\\
\end{tabular}
\end{table}

The nine sources
can be used to explore the low end of the relation between luminosity 
and  burst properties. The long waiting  times between type\,I 
bursts, compared to brighter burst sources, plus the low persistent
emission level make these sources difficult to discover. 
Its large field of view makes the WFC an efficient instrument for
the detection of such rare events.

In Sect. 2 we describe the Chandra observations and
data analysis and in Sect. 3 we discuss which of the detected
sources are the most likely candidates for each burster. In Sect. 4
we briefly present unpublished but relevant observations with
other instruments of \SAXIV\ and GRS\,1741.9$-$2853. In Sect. 5 we
discuss the implications for the class of low persistent emission
bursters.

\section{Observations and data analysis}

With the Chandra satellite (Weisskopf 1988) we observed the WFC error
circles of the five burst sources without persistent emission listed
in Table\,\ref{overzicht}. For each field we used the ACIS-S3 detector
in imaging mode.  We analyzed the level 2 FITS data provided with the
standard data products using the Chandra Interactive Analysis of
Observations Software (CIAO) version 2.1.3. None of the five
observations showed periods of high background and we used all data.
For source detection we used a wavelet-based algorithm (Freeman et~
al. 2002), only taking into account the events between 0.5 and 7
keV. We set the significance threshold for the source detection at
$10^{-6}$, i.e. giving at most one spurious source on the ACIS-S3
detector per observation. In the dithered detector image we checked
each source region for the presence of flickering pixels. If a pixel
detected more than one photon from a source during the whole
observation we marked this as a flickering pixel; given the small
number of photons in each source (see Table\,\ref{result}) we think
that the chance probability that this happens is too small
($\simeq10^{-5}$) to be coincidence. In Table\,\ref{result} we have
noted the source which is affected by a flickering pixel with an
f. The count rate and position are not reliable for this source.  The
wavelet method also gives an estimate of the background. We consider
all sources detected with a significance of more than 3$\sigma$. Only
for the observation of SAX\,J1753.5-2349 no sources above 3$\sigma$
were detected; here we derive an upper limit of 5 counts.

\begin{table}
\caption{Observation log of the bursters at low persistent
emission. For each source we list the start and exposure time of the
Chandra observation, the WFC error radius (99\% confidence), the
absorption column ($N_{\rm H}$, in $10^{21}$ $\cmsq$) as found by
interpolating the HI maps of Dickey \& Lockman (1990), and the
upper limit to the distance ($d_{\rm u}$, in kpc) derived from the
burst peak flux.  For comparison with a model spectrum of a neutron
star H-atmosphere plus power-law we also list for this model the
absorbed flux ($F$, 0.5-7 keV, in $10^{-12}$ $\ergcms$)
corresponding to 1 Chandra count per second, and the absorbed
softness-ratio (SR) of the soft count rate (0.5-2 keV) to the
total count rate (0.5-7 keV). References: a.\ Cornelisse et~al.\
(2002), b.\ Cocchi et~al.\ (2001), c.\ in 't Zand et~al.\ (1998), d.\
this paper
\label{overzicht}}
\begin{tabular}{c@{\hspace{0.25cm}}c@{\hspace{0.25cm}}c@{\hspace{0.25cm}}c@{\hspace{0.25cm}}c@{\hspace{0.25cm}}c@{\hspace{0.25cm}}c@{\hspace{0.25cm}}c@{\hspace{0.25cm}}c@{\hspace{0.25cm}}}
\hline
source   & start date & exp. & $\delta$ & $N_{\rm H}$ & $d_{\rm u}$ & $F$ & SR & ref\\
(SAX\,J) & (MJD)      & (s)  &   ($'$)       &     &           &     &    & \\
\hline 
1324.5$-$6313 & 52162.39 & 5101 & 1.8 & 15  & 6.2 &1.0& 0.69 & a\\
1752.3$-$3128 & 52174.20 & 4717 & 2.9 & 5.6 & 9.2 &2.5& 0.82 & b\\
1753.5$-$2349 & 52187.83 & 5171 & 2.5 & 8.3 & 8.8 &1.6& 0.76 & c\\
1806.5$-$2215 & 52206.39 & 4758 & 2.9 & 12  & 8.0 &1.5& 0.68 & d\\
1818.7$+$1424 & 52092.17 & 4758 & 2.9 & 1.0 & 9.4 &5.5& 0.94 & a\\
\hline
\end{tabular}
\end{table}

\begin{table}[ht]
\caption{For each source detected on the S3-chip we list the position, and
the counts in the total (0.5-7 keV), and the soft (0.5-2 keV) band
as well as the detection significance, $\sigma$.
A conservative estimate for the error in the positions is $0\farcs7$. 
In the second column we indicate -- where appropriate -- reasons to reject 
the source as the burster candidate: o. position outside the WFC
error circle, *. optical counterpart too bright, s. X-ray spectrum too hard
(for discussion see text). In the second column we have also 
indicated the sources which are disturbed by a flickering pixel with
f.
\label{result}}
\begin{tabular}{l@{\hspace{0.2cm}}c@{\hspace{0.2cm}}c@{\hspace{0.2cm}}c@{\hspace{0.2cm}}c@{\hspace{0.2cm}}c@{\hspace{0.2cm}}c@{\hspace{0.2cm}}}
\hline
\# & note & RA & Dec & counts & soft & $\sigma$\\
& & (J2000) & (J2000) & & &\\
\hline
\multicolumn{7}{l}{\bf SAX\,J1324.5$-$6313}\\ 
A &  & $13^{\rm h} 24^{\rm m} 30.2^{\rm s}$& $-63^\circ 12' 41''$& 5.9$\pm$2.4  &  5.9$\pm$2.4 & 3.1\\
B & s & $13^{\rm h} 24^{\rm m} 30.3^{\rm s}$& $-63^\circ 13' 50''$& 77.9$\pm$8.9 & 21.7$\pm$4.7 & 33 \\
C &  & $13^{\rm h} 24^{\rm m} 38.0^{\rm s}$& $-63^\circ 12' 26''$& 6.8$\pm$2.6  &  6.8$\pm$2.6 & 3.4\\
D &  & $13^{\rm h} 24^{\rm m} 38.3^{\rm s}$& $-63^\circ 13' 28''$& 19.5$\pm$4.5 & 19.5$\pm$4.5 & 9.2\\
E &  & $13^{\rm h} 24^{\rm m} 39.4^{\rm s}$& $-63^\circ 13' 34''$& 5.9$\pm$2.4 & 5.9$\pm$2.4 & 3.1\\
\hline
\multicolumn{7}{l}{\bf SAX\,J1752.3$-$3128}\\
A &  & $17^{\rm h} 52^{\rm m} 16.7^{\rm s}$& $-31^\circ 39' 46''$& 10.7$\pm$3.3 & 10.7$\pm$3.3 & 5.3\\
B &  & $17^{\rm h} 52^{\rm m} 30.6^{\rm s}$& $-31^\circ 38' 58''$& 6.7$\pm$2.6 & 4.8$\pm$2.2 & 3.3 \\
C & o & $17^{\rm h} 52^{\rm m} 39.4^{\rm s}$& $-31^\circ 37' 56''$& 36.5$\pm$6.2 & 35.9$\pm$6.1 & 13 \\
\hline
\multicolumn{7}{l}{\bf SAX\,J1806.5$-$2215}\\ 
A & o & $18^{\rm h} 06^{\rm m} 18.1^{\rm s}$& $-22^\circ 15' 39''$& 13.7$\pm$3.9 &  2.8$\pm$1.7 & 5.7\\
B & o & $18^{\rm h} 06^{\rm m} 18.5^{\rm s}$& $-22^\circ 17' 24''$& 48.4$\pm$7.1 & 43.7$\pm$6.7 & 19\\
C & o & $18^{\rm h} 06^{\rm m} 19.9^{\rm s}$& $-22^\circ 18' 03''$&  7.9$\pm$2.8 &  1.9$\pm$1.4 & 4.1\\
D &   & $18^{\rm h} 06^{\rm m} 31.7^{\rm s}$& $-22^\circ 13' 19''$&  9.0$\pm$3.2 &  5.5$\pm$2.4 & 3.9\\
E & * & $18^{\rm h} 06^{\rm m} 35.8^{\rm s}$& $-22^\circ 15' 01''$& 10.5$\pm$3.3 &  9.8$\pm$3.2 & 5.0\\
F & s & $18^{\rm h} 06^{\rm m} 36.8^{\rm s}$& $-22^\circ 15' 26''$& 14.9$\pm$3.9 & $<0.66$      & 7.7\\
G & f & $18^{\rm h} 06^{\rm m} 37.4^{\rm s}$& $-22^\circ 17' 22''$&  7.9$\pm$2.8 &  6.9$\pm$2.6 & 4.0\\
H &  & $18^{\rm h} 06^{\rm m} 43.6^{\rm s}$& $-22^\circ 16' 06''$&  8.7$\pm$3.0 &  8.7$\pm$3.0 & 4.3\\
I & o & $18^{\rm h} 06^{\rm m} 43.8^{\rm s}$& $-22^\circ 18' 42''$& 13.8$\pm$3.7 &  3.0$\pm$1.7 & 6.9\\ 
\hline
\multicolumn{7}{l}{\bf SAX\,J1818.7$+$1424}\\
A & o & $18^{\rm h} 18^{\rm m} 32.1^{\rm s}$& $+14^\circ 22' 09''$& 45.3$\pm$6.9 & 45.3$\pm$6.8 & 21\\
B & o & $18^{\rm h} 18^{\rm m} 34.3^{\rm s}$& $+14^\circ 26' 29''$&  6.8$\pm$2.6 &  4.0$\pm$2.0 & 3.5\\
C & * & $18^{\rm h} 18^{\rm m} 35.6^{\rm s}$& $+14^\circ 22' 32''$&  9.9$\pm$3.2 &  9.9$\pm$3.2 & 5.2\\ 
D & * & $18^{\rm h} 18^{\rm m} 37.6^{\rm s}$& $+14^\circ 22' 44''$& 36.6$\pm$6.1 & 36.6$\pm$6.1 & 18 \\
E &  & $18^{\rm h} 18^{\rm m} 37.8^{\rm s}$& $+14^\circ 22' 06''$&  7.0$\pm$2.6 &  6.0$\pm$2.4 & 3.7\\
F &  & $18^{\rm h} 18^{\rm m} 38.6^{\rm s}$& $+14^\circ 22' 59''$& 27.8$\pm$5.3 & 21.9$\pm$4.7 & 14 \\
G &  & $18^{\rm h} 18^{\rm m} 48.3^{\rm s}$& $+14^\circ 22' 43''$&  9.9$\pm$3.2 &  9.9$\pm$3.2 & 5.1\\
H & o & $18^{\rm h} 18^{\rm m} 55.8^{\rm s}$& $+14^\circ 27' 38''$& 10.2$\pm$3.5 &  7.8$\pm$3.0 & 4.0\\
\hline 
\end{tabular}
\end{table}

\section{Selecting candidate burst sources}

In Table\,\ref{result} we list the detected sources on the whole
S3-chip for each observation, because there is still a 1\% possibility
that the source is outside the error circle. In all four observations
there is more than one source inside or close to the WFC error
circles. Based on the photon count rate, there are no extreme examples
of sources which would qualify them as particularly likely
candidates. This could very well mean that all detected sources
are spurious and none are the bursters, also given that no source
was detected during the observation of \SAXIII.

Thus, we resort to 
several criteria to select viable burster candidates. We
start with excluding all sources outside the WFC error circles.
In Fig.1 we see that several X-ray sources are close to
optical sources from the Sloan Digitized Sky Survey. We have listed
the closest star from the USNO catalogue within 4$''$ of the
X-ray sources in Table\,\ref{usno}.  Given the small number of counts
we estimate an error in the X-ray position of 1 pixel,
i.e. $0\farcs5$, and a systematic error of another pixel. This gives a
total error of $0\farcs7$. The positional error for the sources in the
USNO Digitized Sky Survey is negligible in comparison. For each
source in Table\,\ref{usno} we count the number of stars from the USNO
catalogue inside the WFC error-circle and brighter than the potential
counterpart. On the basis of this number we estimate the chance
coincidence as in the following example.
In the USNO catalogue the star closest to source D of \SAXI\
is at $0\farcs83$ (see Table\,\ref{usno}). We find 24 stars in the
USNO catalogue inside the WFC error circle of $1\farcm8$ that are
brighter than $B$$=$17.2 and $R$$=$15.3. This gives a chance
probability of 0.14\% that {\em one} arbitrarily chosen position
falls within one of the 24 error circles of $0\farcs83$. The five
Chandra sources inside the WFC error circle correspond to five trials, 
i.e.\ the chance probability
that one or more of these sources are close to a star 
is 0.71\%.  In Table\,\ref{usno} we have listed the
chance probabilities $P$ thus computed for all Chandra sources within 4$''$
of an optical star. 

Although it is difficult to draw firm conclusions from a posteriori 
statistics, we think that the optical counterparts with $P<0.1$\%\
are secure. The optical sources located at a distance $\Delta\gtap1''$ 
from the X-ray position are very likely chance coincidences. This 
leaves only D of the \SAXI\ field as a borderline case, which may
or may not be the counterpart.
We have indicated the
optically identified sources in Table\,\ref{result} with an asterisk.

 Assuming that the $V$ magnitudes are between the $B$ and $R$
magnitudes given in the USNO Digitized Sky Survey, we find that the
X-ray to optical flux ratios of these stars are well within the range
of the coronal emission from normal nearby stars found in the ROSAT
All Sky Survey (H\"unsch et~al. 1999; here we use that for coronal
sources with interstellar absorption columns up to $10^{21}$ $\cmsq$
the ROSAT/PSPC count rate is typically 1/3 to 1/4 of the Chandra count
rate).
Soft X-ray transients with neutron stars have X-ray to 
optical flux ratios in quiescence
several orders of magnitude higher than ordinary stars 
(e.g. Fig. 5 in Pooley et~al. 2002a). Thus we conclude
that the optically identified Chandra sources indicated 
in Table\,\ref{result}
are too bright in the optical for them to be the bursters.

\begin{table}
\caption{USNO Digitized Sky Survey sources close to detected Chandra
detections. For each source we give the position, $B$ and $R$ magnitudes, 
the distance ($\Delta$) from the X-ray source, and the chance probability 
$P$ that the optical source is in the Chandra error-circle. 
\label{usno}}
\begin{tabular}{ccc@{\hspace{0.2cm}}c@{\hspace{0.2cm}}c@{\hspace{0.2cm}}c@{\hspace{0.2cm}}c@{\hspace{0.2cm}}}
\hline
\# & RA & Dec & $B$ & $R$ & $\Delta$ ($''$) & $P$ (\%)\\
\hline
D & $13^{\rm h} 24^{\rm m} 38.284^{\rm s}$ & $-63^\circ 13' 27.16''$ & 17.2& 15.3& 0.83 & 0.7\\
\hline
A & $17^{\rm h} 52^{\rm m} 16.723^{\rm s}$ & $-31^\circ 39' 44.24''$ & 19.5 & 17.3& 1.78 & 5.7\\
B & $17^{\rm h} 52^{\rm m} 30.537^{\rm s}$ & $-31^\circ 38' 58.88''$ & 19.0 & 17.3& 1.19 & 1.9\\
\hline
D & $18^{\rm h} 06^{\rm m} 31.589^{\rm s}$ & $-22^\circ 13' 18.94''$ & 15.7& 13.1& 3.32 & 1.1\\
E & $18^{\rm h} 06^{\rm m} 35.819^{\rm s}$ & $-22^\circ 15' 00.87''$ & 16.1& 14.8& 0.28 & 0.0\\
H & $18^{\rm h} 06^{\rm m} 43.734^{\rm s}$ & $-22^\circ 16' 06.08''$ & 19.8& 17.0& 1.86 & 6.3\\
\hline
C & $18^{\rm h} 18^{\rm m} 35.593^{\rm s}$ & $+14^\circ 22' 32.67''$ &  -  & 11.1& 0.66 & 0.0\\ 
D$^a$ & $18^{\rm h} 18^{\rm m} 37.634^{\rm s}$ & $+14^\circ 22' 44.44''$ & 8.6 & 7.4& 0.0 & 0.0\\ 
\hline
\multicolumn{6}{l}{$^a$ star HD\,168344 with $V$$=$7.6}\\
\end{tabular}
\end{table}

Due to the small number of counts in each source, it is not possible
to constrain the spectral shape of the X-ray emission. The most
commonly used models to describe the quiescent emission of neutron
star X-ray transients are 0.3 keV black body radiation, 0.3 keV
Raymond-Smith emission, power-law emission with a slope of
$\Gamma\sim3$ or emission from a hydrogen atmosphere (e.g Campana
et~al. 1998). Added to these models is a hard energy tail detected at
high energies (e.g. Campana et~al. 1998, Asai et~al. 1996). Here we
assume emission from a hydrogen atmosphere of a neutron star (Zavlin
et~al. 1996) plus a power-law, as was, for example, found for the
quiescent emission of the neutron star X-ray transients Cen\,X-4 and
Aql\,X-1 (Rutledge et al. 2001a, 2001b). We
estimate the number of photons below 2 keV using the average
parameters found for Cen\,X-4 and Aql\,X-1, i.e. a power-law
photon-index $\Gamma$=1.0, neutron star radius $R_\infty$=16 km, a
neutron star temperature $kT$=100 eV, and a ratio of the unabsorbed
flux (0.1-7 keV) expected from the H-atmosphere to the power-law
component of 5:1.  For such a spectrum we compute the flux for a
source with 1 Chandra count per second, and the ratio of counts at
energies between 0.5 and 2 keV to the total count rate. The resulting
numbers are listed in Table\,\ref{overzicht}. By comparison with the
observed soft-to-total count ratios listed in Table\,\ref{result} we
can exclude sources which are harder than expected for soft X-ray
transients in quiescence. We do not reject sources with soft spectra,
because there is evidence that the photon-index of the power-law
component could be higher than we have assumed (Campana et~al. 1998).
The two sources which are excluded in this way are indicated in
Table\,\ref{result} with s. Unacceptably high column densities are
needed ($\simeq10^{23}$ $\cmsq$) to account for the lack of soft
photons due to absorption (for source F of \SAXIV\ we do not detect
anything below 2 keV). The fluxes do not
significantly change if we assume the other spectral models or change
the temperature and radius of the neutron star atmosphere model.


For all sources the distribution of the arrival times of the photons 
is compatible with a constant flux. 
Given the small number of photons for each source 
the limits on variability are not very constraining, but we can
exclude that the flux measured is due to a flare lasting shorter
than the exposure time.

Taking all these criteria into account we conclude that
we have four candidate counterparts left for \SAXI\ 
(source A, C, D and E), 
two for \SAXII\ (A and B), two for \SAXIV\ (D and H),
and three for \SAXV\ (E, F and G). For \SAXIII\ we 
have no candidate.

\section{\SAXIV\ and \GRS}

In 't Zand et~al. (1998) reported the detection of two X-ray
bursts from \SAXIV, and showed the analysis of the first burst. 
We take the opportunity of the present paper to report the 
detection of two additional type\,I bursts from \SAXIV\ during 
WFC observations on MJD 50537.91 and MJD 50732.90. No persistent 
emission is observed for this source in any WFC observation.
The first and strongest burst (on MJD 50325.88) has a duration 
of $\simeq$150 s, all other bursts last $\simeq$20 s. The observed 
spectra can be well described by an absorbed black body model with 
temperatures between 1.7 and 2.2 keV. The unabsorbed bolometric peak 
flux of the strongest burst is $(2.6\pm1.2)\times10^{-8}$ $\ergcms$. 
This gives an upper-limit on the distance of 8.0 kpc and neutron star 
radii between 4.8 and 7.0 km, assuming that the peak flux is below the 
Eddington limit of $L_{\rm Edd}$=$2\times10^{38}$ $\ergs$. The waiting
times between the four bursts are 41, 171 and 195 days, respectively. 

Recently the ASM lightcurve of \SAXIV\  became available. It shows
a  faint but clear detection between March 1996 and October 1997. This
coincides with the same period as the occurence of the four X-ray
bursts observed with the WFC.  The maximum persistent flux was
$\simeq2\times10^{-10}$ $\ergcms$ (2-10 keV) , and slowly decreased over time.
This is comparable to the upper-limit derived during the WFC observations.  
Assuming the distance derived above, this corresponds to a luminosity of 
$\simeq2\times10^{36}$ $\ergs$.

\GRS\ was in the field of view of a 47.2 ks ROSAT/PSPC pointed
observation of the Galactic center region obtained on March 2--9,
1992. It is not detected, and we determine an upper-limit of 0.0003
cts s$^{-1}$ (channels 50--240), corresponding to an unabsorbed
luminosity in the 0.5 to 2.5 keV range of $3\times10^{34}$ $\ergs$
at the distance of 7.2 kpc, for an assumed power-law spectrum with
photon index 1 absorbed by a column $\nh=10^{23}$ $\cmsq$ (see Cocchi
et al. 1999). This proves that \GRS\ is a burster with low persistent
emission.

\section{Discussion}

Three of the nine burst sources with low persistent emission
discovered with the WFC were observed during ROSAT observations at
luminosities of $\simeq$$10^{34-35}$ $\ergs$ a few years prior to the
X-ray burst (Kaptein et~al. 2000; Cornelisse et~al. 2002; this
paper). If the five burst sources of Table~\,\ref{overzicht} had
similar luminosities and spectra, their countrate with Chandra would
be several orders of magnitude higher than the countrates of the
sources listed in Table\,\ref{result}.  Instead, the luminosities of
the burst sources are at $\simeq10^{33}$ $\ergs$, comparable to the
BeppoSAX/NFI observations of SAX\,J2224.9$+$5421 (Antonelli
et~al. 1999; in 't Zand 2002, in preparation) and in the range of
quiescent soft X-ray transients with neutron stars (e.g. Campana
et~al. 1998). 

With the interstellar hydrogen column and upper limits to the distances 
listed in Table\,\ref{overzicht} and the
spectrum described in Sect. 3, we compute the unabsorbed flux and upper 
limits to the luminosities between 0.5 and 7.0 keV. For the
brightest candidate counterparts of \SAXI, \SAXII, \SAXIV\ and \SAXV\
we obtain upper limits to the unabsorbed persistent luminosity of
4$\times$10$^{32}$, 3$\times$10$^{32}$, 2$\times$10$^{32}$ and
4$\times$10$^{32}$ $\ergs$, respectively. From the upper-limit
derived from the observation of \SAXIII\ we get a luminosity of
$<4\times10^{32}$ $\ergs$ (0.5-7 keV). These luminosities are indeed
in the range expected for quiescent soft X-transients with a neutron
star.

Because our Chandra observations give more than one possible
counterpart, several possible counterparts must be chance
coincidences. This is in agreement with known $\log N-\log S$
distributions, which predict $\simeq$5 sources in the field of view
(see e.g. Rosati et~al. 2002). This raises the question whether {\em
all} Chandra sources are chance coincidences, i.e.\ whether we have
not detected the actual counterparts for the bursters. Given that
these systems are neutron star low mass X-ray binaries, we compare them
to known other systems, i.e. the soft X-ray transients in quiescence.  
The lowest X-ray luminosities detected for quiescent soft X-ray transients are
$\sim10^{32}$ $\ergs$ (e.g. Cen X-4, Campana et al.\ 1998). This
is around the detection limit for the Chandra observations discussed
in this paper. We therefore consider it possible that we
actually {\em have} detected the persistent flux of the bursters, and
that they are soft X-ray transients in quiescence, for
which no outburst has as yet been detected. If so, this
implies that their actual distances are not much less than the upper
limits listed in Table\,\ref{overzicht}.

The persistent luminosities of the bursters observed with Chandra is
well below the limit set with the WFC observations. This means that we
cannot exclude that the persistent luminosity during the WFC
observations was $\sim$10-100 times higher than detected with Chandra,
and that it was this higher flux level which triggered the burst.  The
detections with ROSAT of 1RXS\,J171824.2$-$402934 and
SAX\,J1828.5$-$1037, and of GRS\,1741.9$-$2853 with GRANAT combined
with non-detections at other epochs, show that the persistent flux
level of these sources is variable.

The energy released during a burst due to nuclear fusion is about 1\%
of the accretion energy of the matter accreted onto the neutron star
(see e.g. Lewin et al. 1993). Dividing the fluence of the bursts
detected with the WFC by 1\% of the persistent emission detected by
Chandra we estimate burst intervals of $\sim$10 years.  If only 1/6th
of the persistent flux is due to accretion, the remainder being due to
the cooling of the neutron star (i.e.  if only the power-law component
is due to accretion, see Sect. 3) the estimated burst intervals rise
to $\sim$60 years. It is also suggested that the power-law
component during quiescence is not due to accretion (see e.g. Campana
et~al.  1998), and this means that the waiting time derived above is
an under-limit. This explains why these events are so rare, and why we have
only seen one burst for most of these sources. 

This raises the question how many of these burst sources with low
persistent emission exist in our Galaxy. With the WFC the
Galactic Center region is observed every half year since 1996, for a
total observation time of $5.5\times10^6$ s up to end 2001. If we
assume the Galactic distribution of low-mass X-ray binaries derived by
van Paradijs \& White (1995), $\simeq$50\% of the population is in the
field of view (40$^\circ$$\times$40$^\circ$) of the WFC. During
all Galactic center observations 5 bursters with low persistent emission 
have been detected, i.e. \SAXII, \SAXIII, \SAXIV,
1RXS\,J171824.2$-$402934, and GRS\,1741.9$-$2853 (the other four are
outside the Galactic center region). This gives an average waiting
time between the detection of these bursters of $1.1\times10^6$ s. If
we also assume that the waiting time between burst of one source is 60
years ($1.9\times10^9$ s) we expect $\simeq2\times10^3$ sources in the
Galactic center region, giving $4\times10^3$ sources in the whole
Galaxy. If on the other hand these sources are extensive periods
of time at a persistent luminosity of $10^{34}$ $\ergs$, as the
detections of SAX\,J1828.5$-$1037, 1RXS\,J171824,2$-$402934, and \GRS\
suggests, the waiting time drops to 0.5 year (see Table\,\ref{previous}).
This gives a number of 30 sources
in our Galaxy. We conclude that the estimates for the total number of
X-ray bursters with low persistent fluxes range from 0.5 to 60 times
the number of known bursters ($\simeq70$).

In this respect it is interesting to note that the first Chandra observations
of globular clusters indicate that these systems harbour more quiescent soft
X-ray transients than bursters with high persistent fluxes. For example, 
Liller\,1, NGC\,6440 and NGC\,6652 all contain such quiescent sources in 
addition to
the bright source (Homer et al.\ 2001, Pooley et al.\ 2002b, Heinke et al.
2001); and 47\,Tuc, $\omega$\,Cen, NGC\,6752 and NGC\,6397 contain quiescent
sources but no bright source (Grindlay et al.\ 2001a, 2001b, Rutledge et
al.\ 2001c, Pooley et al.\ 2002a).
The formation mechanism for low-mass X-ray binaries 
in globular clusters (tidal capture or exchange encounter; see
review by Hut et al.\ 1992) is different from the formation mechanism
in the galactic disk (evolution of a primordial binary). 
If the ratio of quiescent to bright X-ray bursters depends on
the formation mechanism, we do not necessarily expect comparable ratios in
the cluster and in the Galactic disk.

If bursts can arise from quiescent systems, we must consider the
possibility that a burst from a globular cluster is due to a dim source, 
rather than to the bright source in it. This would undermine the
argument that a burst from a cluster proves that the bright source 
in it is a neutron star. Nonetheless, we think that the
argument holds in all eleven cases where it has been applied so far,
as bursts from dim sources are extremely rare. For example,
we have detected $\sim2200$ X-ray bursts in our WFC observations
of the Galactic center region; only five of these are from dim sources.
Indeed, bursts from the globular cluster NGC\,6440 were detected
only when the transient in this cluster was active 
(in 't Zand et~al. 2001).

\end{document}